\documentclass[default]{aastex7}
\usepackage{amsmath}
\usepackage{caption}

\newcommand{\beq}{\begin{equation}}
\newcommand{\eeq}{\end{equation}}
\newcommand{\AU}{{\rm AU}}
\newcommand{\St}{{\rm St}} 
\newcommand{\Ma}{{\rm Ma}} 
\newcommand{\Rey}{{\rm Re}} 
\newcommand{\Kn}{{\rm Kn}} 

\begin{document}

\title{Strong Toroidal Recirculation Zones at the Inner Dust Rim 
as the Origin of Meteoric Chondrules and Calcium-Aluminum Inclusions}

\author[orcid=0000-0002-7369-6235,sname='Williams']{Peter Todd Williams}
\affiliation{The Aerospace Corporation}
\email[show]{peter.todd.williams@gmail.com}  

\begin{abstract}
We hypothesize strong (transonic) twin toroidal recirculation zones above and below the accretion disk midplane, 
rather close-in to the protosun, to be the source of chondrules and calcium-aluminum inclusions (CAIs).
The recirculation zones act as centrifugal separators. In the case of
chondrules, we suggest this happens during
Class II (T~Tauri) stage of protostellar accretion, and in the case of CAIs, during an earlier higher-$\dot M$ phase of accretion. 
The recirculation zones advect and raise dust and solid aggregates above the midplane, making a ``mushroom-cap,'' and they also
 generate weak standing oblique shocks that heat and fuse protochondrules.
We do not model the emission of these shocks, but point out that they will produce Doppler-broadened,
possibly twin-peaked line emission with width of the order of $\simeq 200\ {\rm km\ s^{-1}}$.

For concreteness, we focus on chondrules in the paper.
Small ($\lessapprox 10 {\rm\ \mu m}$) diameter protochondrules are evaporated by the standing shocks, whereas large ($\gtrapprox 1\ {\rm cm}$) protochondrules are
too heavy to be entrained and accelerated by the outer recirculation zone and outflow.
Intermediate-size protochondrules, however, are centrifugally ejected and carried
by high-speed diffuse gas outflow to the outer regions of the disk, where they rain down. 

The recirculation-induced dust mushroom caps will create significant IR continuum emission. We suggest they
 coincide with observed inner ``puffed-up'' dust rims.
We suggest that the recent interferometric inferences of arcs or ellipses in the sub-AU continuum IR of low-mass Class II protostars 
may be observation of corresponding chondrule-producing recirculation zones in those systems.
\end{abstract}



\section{Introduction} \label{sec:introduction}

Chondrules are spherical inclusions, typically a millimeter or so in diameter, which along
with calcium-aluminum  inclusions (CAIs) are contained in most chondrites, the latter being the stony non-metallic bodies that form the majority of meteorites that fall to Earth's surface. 
Chondrules and CAIs date to 4.567 Gyr, with CAIs being formed first,
and chondrules most likely following 2-3 Myr later.
It is as yet unknown how either came to be.
A large and diverse set of suggestions regarding their origin have been proposed in the literature. A concise
discussion of the relative merits of the better-known hypotheses regarding chondrules in particular
 is provided by \citet{Boss_1996}, and a fairly recent review of the problem of chondrules may be found in \citet{ConnolyJones_2016}.

A theory of chondrules must satisfy a large number of constraints, not all of which we will address here.
See \citet{Wasson_1993}. We address
(a) their narrow range in ages (a spread
 of at most $\pm$ 1-2 Myr for chondrules, and even less for CAIs), 
(b) their narrow range in size and mass, 
(c) evidence of heating with peak temperatures lasting for minutes to hours at most,
(d) evidence for rapid (eg 0.5 - 400 K/hr, depending on composition) cooling, (e) evidence for some if not all chondrules undergoing
multiple episodes of heating, (f) presence of chondrules in parent bodies beyond $1\ {\rm AU}$. We believe the model described here is also consistent with additional constraints ({\em e.g.} high ambient partial pressure of volatiles, magnitude
of inferred magnetic field upon cooling to Curie point, etc), but we do not elaborate on these constraints here.

We will focus on chondrules in the development that follows, but we expect that aspects of the model can be made to apply to
CAIs as well, presenting a unified picture of how both came to be.

\section{Assumed Flow Conditions}

Let us suppose a $1\ {\rm M}_\Sun$ protostar, and let us suppose an accretion disk extending at least down to $0.1~\AU$, somewhat
outside of where
we imagine the dust evaporation radius to be. (Gas-phase accretion may or may not continue beyond this point.)

We will describe the following notional zones: (1) the dusty disk, which terminates in a ``mushroom'' cap (as seen in meridional
section) or ``puffed-up'' inner disk rim, (2) inner twin toroidal recirculation zones (iTTRZ), above and below the disk, set slightly
back ({\em i.e.} behind) the mushroom cap, (3) standing shocks in the iTTRZs, where the protochondrule fusion occurs, (4) outer 
twin toroidal recirculation zones (oTTRZs) that are hotter, faster and more rarefied than the iTTRZs, (5) radially-outwards stream flows
(SFs) even farther out, emanating from the recirculation zones and mushroom cap, that are even
hotter, faster, and less dense than the oTTRZs, (6) a diffuse bound envelope that is radially farther out still, the 
details of which we do not address other than that we rely upon it to catch the outflowing chondrules to allow them to
rain down upon the outer ($\gg 1\ \AU$) regions of the disk.

At the fiducial radius of $0.1\ \AU$, just outside of where we suppose the mushroom cap inner disk rim to be,
let us suppose the disk to be composed of largely molecular hydrogen, and let it have a midplane density of $10^{-8}\ {\rm g\ cm^{-3}}$.
The mean free path is $\lambda=0.14\ {\rm cm}$.
Note that the Keplerian orbital speed here is just shy of $100\ {\rm km\ s^{-1}}$. If chondrules are formed nearby, as we suppose,
then we will have to accelerate them to speeds of $130\ {\rm km\ s^{-1}}$ if we hope to get them to the outer reaches of the
accretion disk.
Let the disk midplane temperature be $800\ {\rm K}$, and let it have a thickness of $H\simeq c_s/\Omega_{\rm K}$.
Given the column densities, we suppose this region of 
dusty molecular gas to be optically thick to its own blackbody radiation and to the radiation from the protostar.
Despite high UV opacities in this dusty region, shielding it from photoionization,
the temperature is just sufficient to ensure that the magnetorotational instability (MRI) remains active due to thermal ionization
of alkali metals, and so we assume this region to be turbulent.

Let the inner twin toroidal recirculation zones (iTTRZs) be composed of dissociated hydrogen, with a temperature of
$1600\ {\rm K}$, a sound speed of $c_s \approx 5\ {\rm km\ s^{-1}}$, and a density of $10^{-9}\ {\rm g\ cm^{-3}}$.
The gas would be molecular in LTE, but the standing shock dissociates the hydrogen, and because most micron-size
dust has evaporated here, there are no efficient recombination channels, and so the gas remains dissociated.
 Let the 
recirculation zones circulate with a gas speed slightly in excess of the sound speed. The overturn period of the
iTTRZs will be $P \simeq 2\pi/\Omega_K \simeq P_K$, which will be on the order of 300 hours. 
As we will see, micron-size grains will evaporate in the iTTRZs, helping to ensure that, with column densities on the
order of $\Sigma = 10^2\ {\rm g\ cm^{-2}}$, the region is nevertheless optically thin in the optical and mid-IR.
The low supersonic value of this Mach number is important, as it ensures the development of a spiral
oblique shock. Such spiral oblique shocks in transonic vortices are known from the literature.
See, for example, \citet{Zhang_Chen_Li_Jiang_2018}. We will suppose this to be a Mach~$1.5$ shock.

We then suppose the iTTRZs to be enveloped within larger outer twin recirculation zones (oTTRZs). Suppose dissociated hydrogen
with a temperature of $5000\ {\rm K}$ and a density of $10^{-10}\ {\rm g\ cm^{-3}}$. The mean free path 
is $\lambda = 34\ {\rm cm}$ and the sound speed is $c_s = 8\ {\rm km\ s^{-1}}$.
Suppose the Mach number of this outer zone to be higher: let $\Ma = 2.5$ ({\em i.e.} the meridional gas speed is $20\ {\rm km\ s^{-1}}$).
Again, the overturn time of this outer region of the vortices will be of the order of $300\ {\rm hr}$.

Almost finally, in the last shell of this onion-skin model, we imagine stream flows (SFs) that are at most very slightly bound if not unbound: in addition to the Keplerian azimuthal velocity, let them have a meridional ($R-z$) velocity of $85\ {\rm km\ s^{-1}}$, corresponding to
a Mach~5 meridional flow, in $10,000\ {\rm K}$ gas of ionized hydrogen with a density of $10^{-12}\ {\rm g\ cm^{-3}}$,
and a sound speed of $17\ {\rm km\ s^{-1}}$.

And truly finally, we imagine an outer envelope that we do not bother to model other than to suppose it to be far out radially, bound,
and sufficiently dense to catch chondrules that are flung out to the outer reaches, allowing them to rain down upon
the disk's hinterlands. We may suppose similar gas conditions to those in the SF regions mentioned above, but without the 
large meridional velocity.

\section{Candidate Protochondrule Histories}
Let us now follow a set of candidate protochondrules through this environment. Suppose six notional
protochondrules: A, B, C, D, E, and F. Let their masses be such that, were they to fuse into melt spherules, their
diameters would be $1\ {\rm \mu m}$, $10\ {\rm \mu m}$, $100\ {\rm \mu m}$, $1\ {\rm mm}$, $1\ {\rm cm}$, and $10\ {\rm cm}$
respectively. We will show that the flow field as described acts as a natural sieve: it sifts out large particles, which
can not be lifted and entrained into the iTTRZS or accelerated and entrained in the oTTRZs and SFs,
and it also removes small particles, which evaporate. We will thus
have a simple way of explaining the narrow range of sizes and masses of chondrules.

This holds even if we imagine the candidate protochondrules to begin as somewhat ``fluffy'' flocculates. Theoretically,
in diffusion-limited cluster aggregation (DLCA), allowing for cluster rotation, the effective fractal dimension of aggregates
may be as low as $1.55$ \citep{Jungblut_etal_2019}. However, in the inner disk regions we are considering, aggregates will have been collisionally
compacted. Let us suppose a more reasonable effective fractal dimension of $2.5$.

For the purposes of drag, the effective size of a fractal aggregate depends upon Knudsen number $\Kn$. The mean free path in the
disk is small, however, so that for the larger flocculates we are safely in the Stokes regime. The drag can then be based on
the effective diameter $D = D_d N^{1/\delta}$ where $\delta$ is the fractal dimension. Strictly speaking, for the smaller
particles with large $\Kn$, we should use a somewhat smaller diameter corresponding to the projected area \citep{Stoyanovskaya_etal_JoP_2020}. 
However, this does not really
matter here, as the point is to evaluate whether the protochondrules have been diluted by settling, and this will not happen
for the smaller candidate flocculates in any case.

The stopping time in the Stokes regime is 
\beq 
\tau_s^{({\rm S})} = \frac{1}{9} \frac{D^2}{\lambda \bar c} \frac{\rho_s}{\rho_g},
\eeq
whereas the stopping time 
in Epstein regime is
\beq 
\tau_s^{({\rm E})} = \frac{1}{2} \frac{D}{\bar c} \frac{\rho_s}{\rho_g},
\eeq 
where $\bar c$ is the thermal speed, and the effective $\rho_s \simeq 3.25\ {\rm g\ cm^{-3}}$ for a primitive grain, but less for a flocculate.
In general, both stopping times may be evaluated and the stopping time can be taken to be
\beq 
\tau_s = \max(\tau_s^{({\rm E})}, \tau_s^{({\rm S})}).
\eeq
(This can easily be seen to be equivalent to the stopping time given the drag prescription of \citet{Weidenschilling_MNRAS_1977},
for low $\Rey$.)

The Stokes number is $\St = \tau_s \Omega_K$.
The solids, depending on effective density, have long since settled to a scale height $H_s$ related to gas scale height $H_g$ by
\beq 
\frac{H_s}{H_g} = \sqrt{\frac{\alpha}{\alpha + \St}}
\label{eq:scaleheight}
\eeq
\citep{DubrulleMorfillSterzik_1995,Youdin_and_Lithwick_2007} 
where $\alpha$ is the usual Shakura-Sunyaev parameter and we take $\alpha = 0.001$. To be entrained in the iTTRZs, which are 
above and below the midplane at $1/10^{th}$ its density, the aggregates will have to be lofted $N_{sh} \approx 2$ density scale heights. We calculate the retained fractions
$F_r$ where
\beq 
F_r = \exp{(N_{sh} - N_{sh} H_g^2/ H_s^2)}
\label{eq:retained}
\eeq
Note that $F_r \approx 0$ for group F: these large agglomerates, even if we allow them to be
``fluffy'' with fractal dimension $\delta = 2.5$, are simply too heavy and are not lofted far enough above and below the midplane to be 
entrained into the iTTRZs. If we had assumed spherules instead of fluffy aggregates, we also would have filtered out most of 
group E (the nominally $1\ {\rm cm}$ spherules). The $1\ {\rm mm}$ and smaller protochondrules, whether fluffy aggregates or spherules,
remain and can be lofted and entrained into the iTTRZs.
See Table~{\ref{tab:lofting}}.

Once in the iTTRZs, the temperature is hot enough that we must now worry about evaporation. Let us suppose that agglomerates melt here and form 
spherules, an assumption we will justify further below.
The evaporative mass flux per surface area $J$ of a spherule above the solidus temperature is approximated by the Hertz-Knudsen relation.
We have
\beq 
J = \alpha_\nu (P_v-P_{amb}) {\sqrt{\frac{m}{2\pi k_B T_s}}}
\eeq
where $P_{amb}$ is the ambient partial pressure, and the equilibrium vapor pressure $P_v$, is a strong function of $T_s$.

The vapor pressure $P_v(T_s)$ of enstatite may be approximated via the Clausius-Clapeyron equation, which we write
in the empirical form
\beq {}
P_v = 10^6\ {\rm erg\ cm^{-3}} \times \exp{(A-B/T_s)}
\eeq 
using numerical values $A = 20.3$ and $B=65303\ {\rm K}$ \citep{Nagahara_Kushiro_Mysen_1994}.

Thermal diffusivities of chondrules are high enough that we can consider them to be in internal thermal equilibrium.
Temperature may then be found by simple surface flux thermal balance:

\beq 
Q_{\rm coll}(T_g) = Q_{\rm rad}(T_s) + Q_{\rm evap}(T_s).
\eeq
Here, the collisional heating per unit surface area is approximated as 
\beq 
Q_{\rm coll} = \alpha_a n_{\rm H} \bar c \left(k_{\rm B}(T_g-T_s)\right)
\eeq
where $\alpha_a$ is the accommodation coefficient and $\alpha_a \approx 0.1$ is good for silicates, and $n_{\rm H}$ is number density.
Radiative cooling is just the usual $\varepsilon \sigma_{\rm SB} T^4$, and we take the emissivity to be $\varepsilon = 0.9$.
Evaporative cooling is
\beq 
Q_{\rm evap} =  J L_e
\eeq
where $L_e$ is the latent heat of evaporation. We take $L_e \approx 3.86\times 10^{10}\ {\rm erg\ g^{-1}}$.

Because all heating and cooling scales with chondrule surface area, there is a size-independent equilibration temperature. 
For the iTTRZs, this is $1006\ {\rm K}$. 
Even micron-sized grains can survive this temperature for many overturn times. Such grains will
 not survive the assumed ${\Ma}=1.5$ standing oblique shock, however. It is the nature of a recirculation zone that most of the
gas --- and therefore the low-$\St$ particulates as well --- that finds itself there goes around many times. The net result is that
we expect the micron-scale particulates, which form the major contribution to the mid-IR opacity, will have evaporated in the iTTRZs.


\begin{figure} 
\includegraphics[width=6in]{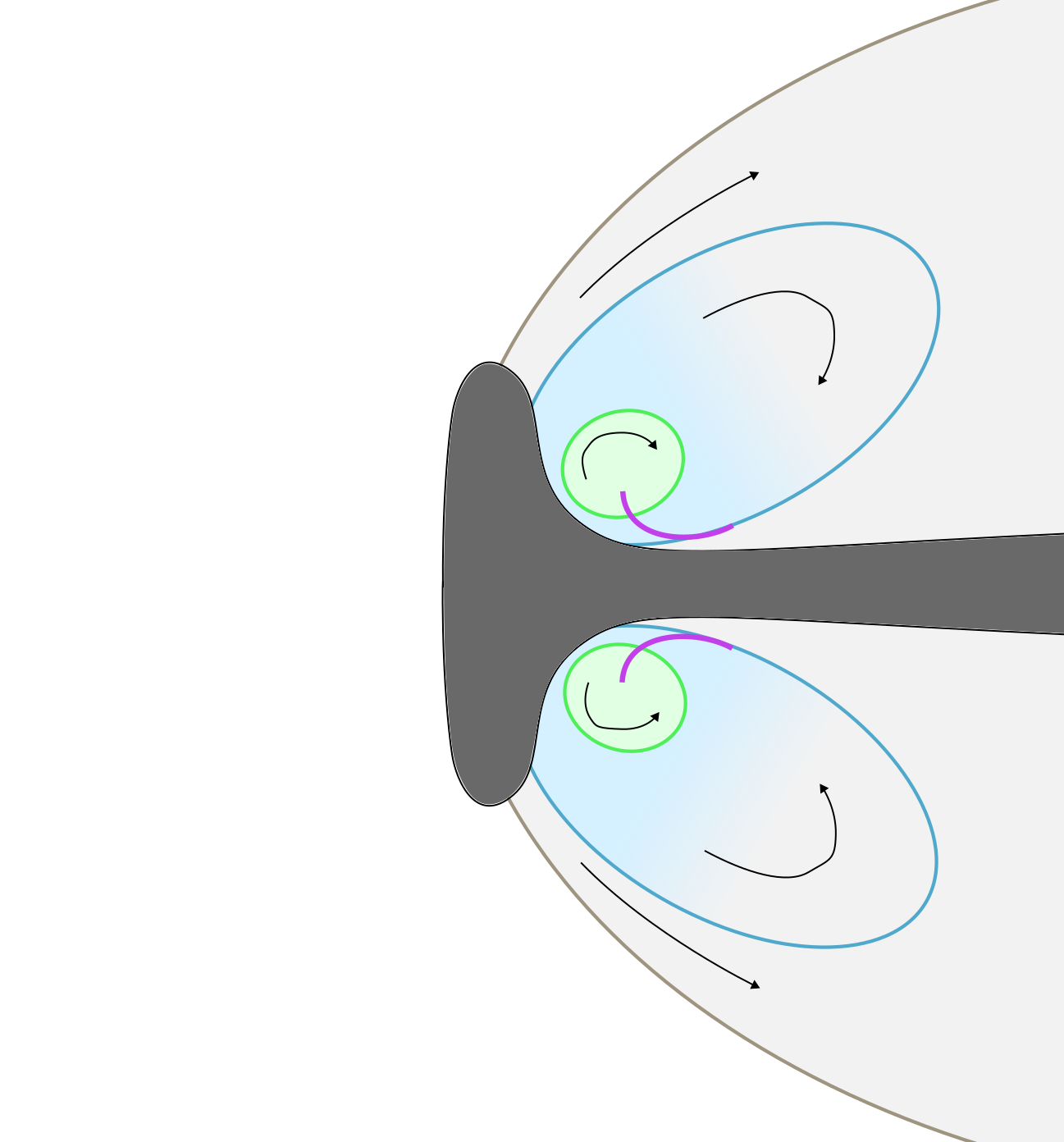}
\captionsetup{width=4in}
\caption{Figure showing meridional section of a 
notional disk with dusty mushroom cap, inner and outer toroidal recirculation zones, and outer stream flows,
the shape of which is superficially suggestive of a bow shock. In this paper, protochondrule heating is supposed to take place
in an oblique recompression shock (mauve) where the recirculation zones impinge upon the disk. Protostar, not shown, is off the
page to the left.}
\label{fig:closeup}
\end{figure}

\begin{deluxetable*}{lrrrrrrr}
\digitalasset
\tablewidth{0pt}
\tablecaption{Settling and Retained Fractions of Aggregates and Spherules \label{tab:lofting}}
\tablehead{
\colhead{Quantity} & \colhead{A} & \colhead{B} & \colhead{C} & \colhead{D} & \colhead{E} & \colhead{F} 
}
\startdata
Notional chondrule diameter (cm) & $1.00 \times 10^{-4}$  & $1.00 \times 10^{-3}$  & $1.00 \times 10^{-2}$  & $0.1$  & $1$  & $10$  \\ 
\hline
diameter of aggregate (cm) & $1.00 \times 10^{-4}$  & $1.58 \times 10^{-3}$  & $0.0251$  & $0.398$  & $6.31$  & $100$  \\
Kn of aggregate  & $1.43 \times 10^{3}$  & $90.4$  & $5.71$  & $0.36$  & $0.0227$  & $1.43 \times 10^{-3}$  \\
Epstein (E) or Stokes (S) & E  & E  & E  & E  & S  & S  \\
stopping time:  & $0.1\ {\rm s}$  & $0.2\ {\rm s}$  & $0.9\ {\rm s}$  & $3.5\ {\rm s}$  & $2.3\ {\rm m}$  & $2.4\ {\rm h}$  \\
Stokes number & $3.51 \times 10^{-7}$  & $1.40 \times 10^{-6}$  & $5.57 \times 10^{-6}$  & $2.22 \times 10^{-5}$  & $8.64 \times 10^{-4}$  & $0.0545$  \\
scale height ratio:  & $1$  & $0.999$  & $0.997$  & $0.989$  & $0.733$  & $0.134$  \\
retained fraction:  & $0.999$  & $0.997$  & $0.987$  & $0.95$  & $0.137$  & $3.19 \times 10^{-55}$  \\
\hline
Kn of spherule  & $1.43 \times 10^{3}$  & $143$  & $14.3$  & $1.43$  & $0.143$  & $0.0143$  \\
Epstein (E) or Stokes (S) & E  & E  & E  & E  & S  & S  \\
stopping time:  & $0.1\ {\rm s}$  & $0.6\ {\rm s}$  & $5.6\ {\rm s}$  & $56.0\ {\rm s}$  & $14.5\ {\rm m}$  & $1.0\ {\rm d}$  \\
Stokes number & $3.51 \times 10^{-7}$  & $3.51 \times 10^{-6}$  & $3.51 \times 10^{-5}$  & $3.51 \times 10^{-4}$  & $5.45 \times 10^{-3}$  & $0.545$  \\
scale height ratio:  & $1$  & $0.998$  & $0.983$  & $0.86$  & $0.394$  & $0.0428$  \\
retained fraction:  & $0.999$  & $0.992$  & $0.922$  & $0.445$  & $3.55 \times 10^{-6}$  & $0$  \\
\enddata
\end{deluxetable*}

\begin{deluxetable*}{lrrrrr}
\digitalasset
\tablewidth{0pt}
\tablecaption{Sorting and Thermal Processing in iTTRZs and Shocks \label{tab:iTTRZ}}
\tablehead{
\colhead{Quantity} & \colhead{A} & \colhead{B} & \colhead{C} & \colhead{D} & \colhead{E} 
}
\startdata
Notional chondrule diameter (cm) & $1.00 \times 10^{-4}$  & $1.00 \times 10^{-3}$  & $1.00 \times 10^{-2}$  & $0.1$  & $1$  \\
\hline
stopping time in iTTRZ:  & $0.3\ {\rm s}$  & $2.8\ {\rm s}$  & $28.0\ {\rm s}$  & $4.7\ {\rm m}$  & $46.7\ {\rm m}$  \\
Stokes number in iTTRZ:  & $1.76 \times 10^{-6}$  & $1.76 \times 10^{-5}$  & $1.76 \times 10^{-4}$  & $1.76 \times 10^{-3}$  & $0.0176$  \\
\hline
stopping time in shock:  & $0.1\ {\rm s}$  & $1.3\ {\rm s}$  & $13.4\ {\rm s}$  & $2.2\ {\rm m}$  & $22.3\ {\rm m}$  \\
evap time post-shock at Mach 0.51:  & $1.0\ {\rm s}$  & $10.5\ {\rm s}$  & $1.7\ {\rm m}$  & $17.5\ {\rm m}$  & $2.9\ {\rm h}$  \\
evap time post-shock at Mach 0.0:  & $5.3\ {\rm s}$  & $52.7\ {\rm s}$  & $8.8\ {\rm m}$  & $1.5\ {\rm h}$  & $14.6\ {\rm h}$  \\
\enddata
\end{deluxetable*}

Heating upon passage through a shock is actually not trivial, because particles radiate, and this affects the thermal structure of the
shock both before and after the shock, possibly leading to a thick, optically-thick shock, preceded and followed by optically-thin regions.
See \citet{Ciesla_and_Hood_2002}. The following simple analysis ignores these complications other than to assume a thick, optically-thick
post-shock region;
we expect that shock conditions can be adjusted accordingly in response to more detailed analysis.

Consider a standard hydrodynamic Mach~$1.5$ shock following Rankine-Hugoniot conditions. 
Relative to the post-shock gas, the particles will initially be moving at Mach~0.51. This introduces additional heating, approximated as
an additional ``ram'' heating term. More significantly, let us suppose that, due to the increased concentration of particles in the
post-shock region, it is optically thick to the particles' own thermal radiation. This means it will also no longer be heated
by illumination from the protostar, but the loss of the radiative cooling channel is far more impactful. 

These particles are all now firmly in the Epstein regime (large $\Kn$), but they are now transonic, introducing additional drag.
\citet{Kwok_1975} constructed a simple way to estimate drag across a range of Mach numbers in this regime. We will discuss this more
when we get to the oTTRZs and the wind. Here however, the effect of this increased drag 
is simply to reduce the stopping times below what we have estimated based on Epstein. In any case, the evaporation times will
still be longer than the stopping times; the result is that we may estimate the evaporation times assuming the particles are
stationary with respect to the post-shock gas. Then we find that if, say, we assume a 
shock thickness of $\approx 400\ {\rm km}$, corresponding to a transit time of just under $\Delta t \simeq 1\ {\rm min}$, then
the $1\ {\rm \mu}$ and $10\ {\rm \mu m}$ will have both evaporated, but the larger particles will remain (at somewhat reduced diameter).
We will keep track of this reduced diameter in a Monte Carlo simulation, but ignore it for the purposes of our tables.

A portion of the particles will be flung out each time they go around the recirculation zone. We estimate very roughly that a particle
will circulate a number of times equal, on average, to $\simeq 1/\St$ before being ejected.
This may be modeled in a MC scheme by comparing $\St$ to a random number on the interval $(0,1)$.

Upon being ejected from the iTTRZs, a sizeable fraction of particles will find themselves in the oTTRZs. These regions are moving much
faster, and so the particles will have relative motion, with an upper limit of roughly the speed of the recirculation of the oTTRZs themselves,
which is Mach~2.5. The particles will of course accelerate to match the gas speed (depending on their Stokes numbers), but in the meantime
we confirm that the additional collisional and ram heating at such speeds are not sufficient to evaporate the particles. Let us say that chondrules with Stokes numbers on the order of $0.1$ or more will not be accelerated, and will be lost.
To estimate $\St$, we must now consider the additional drag due to supersonic motion. A rough estimate for large $\Ma$ is the ballistic
or Newtonian stopping time. Stopping time in Newton ballistic regime is 
\beq 
\tau_s^{({\rm N})} = \frac{2}{3} \frac{D}{u} \frac{\rho_s}{\rho_g} = \frac{2}{3} \frac{D}{c_s \Ma} \frac{\rho_s}{\rho_g}.
\eeq
It is worth noting that the Kwok stopping time \citep{Kwok_1975} is
\beq 
\tau_s^{({\rm K})} = \frac{2}{3} \frac{D}{\sqrt{\frac{16}{9}\bar c^2 + u^2}} \frac{\rho_s}{\rho} =
\left( \left(\tau_s^{({\rm E})}\right)^{-2}+ \left(\tau_s^{({\rm N})}\right)^{-2} \right)^{-1/2},
\eeq
but for rough estimates, the following approximation suffices:
\beq 
\tau_s^{({\rm K})} \simeq \min(\tau_s^{({\rm E})}, \tau_s^{({\rm N})}).
\eeq

In any case, for $\St \ge \St_{\rm crit} \approx 1/10$, we expect particles will not follow gas streamlines, and will be lost.
In the ballistic limit, this implies an upper size cutoff; we require
\beq 
D \le \St_{\rm crit} \frac{3}{2} \frac{c_s \Ma}{\Omega} \frac{\rho_g}{\rho_s}
\eeq
For the oTTRZs, we have $D \lessapprox 1.5\ {\rm cm}$, and for the streams, $D \lessapprox 0.6\ {\rm cm}$. All larger particles
will be filtered by these last stages; they will not be accelerated by the oTTRZs and the streams, and will fail to reach $R > 1\ {\rm AU}$.

\section{Monte Carlo}
We build a simple toy Monte Carlo model to show the effect of the processing stages discussed above. We begin with a million
 candidate protochondrules
that are fluffy (fractal dimension $\delta = 2.5$, built from $1\ {\rm \mu m}$ diameter primitive grains), and another million candidate protochondrules that are spheroidal. The candidate protochondrules are all composed of
a random number of between $1$ and $10^{18}$ (uniform in the log) primitive $1\ {\rm \mu m}$ 
grains of density $3.25\ {\rm g\ cm^{-3}}$.

For each particle, we then determine the drag regime, find the Stokes number, and then the corresponding
retained fraction from eq.~({\ref{eq:retained}}). The calculated 
retained fraction is compared with a random number drawn from the uniform unit distribution,
and particles are correspondingly deleted or retained from the simulation. Particles then move on to thermal 
processing by the iTTRZs and shocks. At this point, all protochondrules are assumed to be spheres. Each time around the iTTRZs, particles
have a chance to exit the vortices, with a probability taken to be $1/\St$ where $\St$ is the re-calculated Stokes number corresponding to the
local conditions. Upon passage through a shock, it is assumed that all protochondrules lose $10\ {\rm \mu m}$ of diameter.

The rationale for this is that, again, the equilibrium temperature is, at least in this simple theory assuming internal thermal equilibrium,
independent of particle diameter. Therefore, assuming all particles experience the same amount of time passing through the
hot, optically-thick portion of the shock, they will lose the same amount of mass per surface area, {\em i.e.} the same loss in diameter.

We now have a scenario which, we argue, satisfies conditions (a) through (g) above. In order: (a) the scenario corresponds neatly to the
duration of the Class II stage of solar accretion, which fits the age and age spread of chondrules; (b) as the Monte Carlo simulation
shows, the scenario can reproduce a suitable narrow range in sizes; (c) and (d) 
like all shock heating models, heating occurs rapidly with peak temperatures lasting several minutes
at most, and cooling times consistent with experimental constraints; (e) almost all chondrules in the MC simulation undergo multiple
episodes of heating; (f) chondrules are given velocities just below escape velocity, and easily reach $1\ \AU$ or greater.


\begin{figure}
\plotone{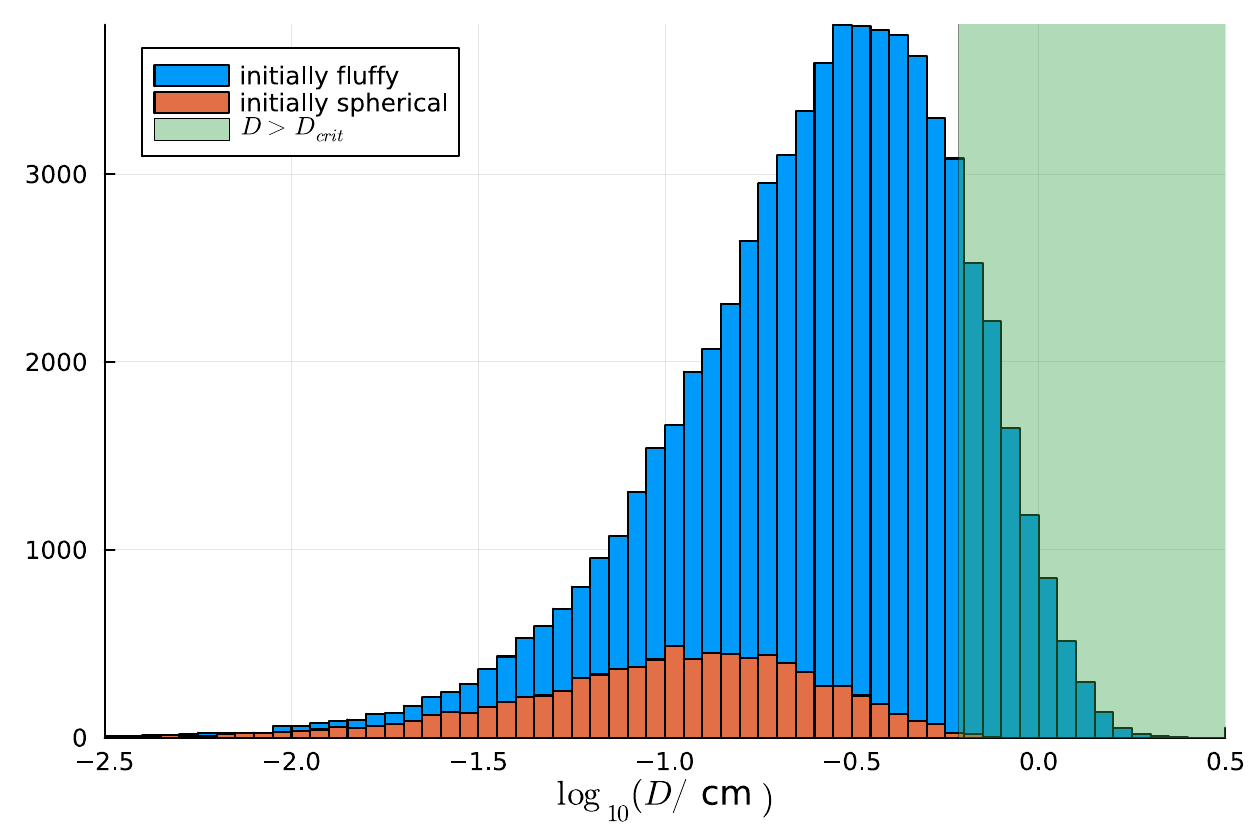}
\caption{Histograms of melt spherule diameters in Monte Carlo model, beginning with two million candidate protochondrules evenly
distributed (in the logarithm) between $10^3$ and $10^{18}$ primitive $1\ {\rm \mu m}$ diameter grains, half starting as fluffy aggregates,
and half starting as spherules. Histogram is left unnormalized to help
illustrate difference in survival fraction of initially fluffy ($6.1\%$ survival) vs. initially spheroidal ($0.8\%$ survival).
Protochondrules in the green-shaded region will not be entrained and dispersed by the streams into the broader disk ($> 1\ \AU$), but this is
not modeled here. A slightly lower choice of gas densities will push these curves further to the left, in closer correspondence
with observations, but we
have done little-to-no tuning here, preferring to work with round numbers for this toy model.}
\label{fig:hist}
\end{figure}

\section{Discussion}
We have presented a simple model of how nested toroidal recirculation zones during the early protostellar accretion phase of the Sun might
explain chondrules (and CAIs). This is a simple model, whereas the shock heating of solid-laden gas is not a trivial calculation. This
bears emphasis, because the rates of evaporation of chondrules (for example)
are exquisitely sensitive to temperature and duration of heating, but these
quantities can not easily be accurately predicted. Our intent here is not to argue in detail that toroidal recirculation zones --- coupled
with outflow streams to carry protochondrules to the outer disk regions --- can explain chondrules and CAIs; our intent is simply to argue
that this is a plausible scenario worthy of further study. We suggest that, in the meantime, toroidal recirculation zones offer one possible scenario that might help explain the enigma of chondrules and CAIs. In particular, the model explains multiple reheating events, and it
explains the narrow range of masses. The latter happens because large particles are filtered at two stages: first, especially if they
are spherules, they are not lifted
above the disk and entrained in the recirculation zones; second, even if, thanks to being ``fluffy'' aggregates, they may be lifted into the
recirculation zones, they can not then be effectively ``flung'' to larger distances from the protostar, as their Stokes number is too high.
Conversely, small protochondrules are lost quite simply because they are evaporated in the recirculation zones.

{}
\bibliography{chondrules_super_short_}{}
\bibliographystyle{aasjournalv7}



\end{document}